# Metal-Insulator Transition in Mn Perovskite Compounds

Yukitoshi Motome[1)], Hiroki Nakano[2)] and Masatoshi Imada[2)],
*1) Department of Physics, Tokyo Institute of Technology, Oh-okayama, Meguro-ku, Tokyo 152-8551, Japan*
*2) Institute for Solid State Physics, University of Tokyo, Roppongi, Minato-ku, Tokyo 106-8666, Japan*
*e-mail address imada@issp.u-tokyo.ac.jp*

**Abstract**
We discuss Mott insulating and metallic phases of a model with $e_g$ orbital degeneracy to understand physics of Mn perovskite compounds. Quantum Monte Carlo and Lanczos diagonalization results are discussed in this model. To reproduce experimental results on charge gap and Jahn-Teller distortions, we show that a synergy between the strong correlation effects and the Jahn-Teller coupling is important. The incoherent charge dynamics and strong charge fluctuations are characteristic of the metallic phase accompanied with critical enhancement of short-ranged orbital correlation near the insulator.

# 1 Introduction

Transition metal oxides with perovskite structure offer many prototypical systems which show various types of strong correlation effects [1]. Mott insulating states are widely observed phenomena in these compounds at integer fillings of $3d$ electrons. A typical property of metals found near this Mott insulating state is that they are "bad metals" where transport and optical measurements show strong dampings and dominance of incoherence in the charge dynamics. The optical conductivity generically has a broad and long tail up to the order of eV, which is in contrast with the prediction of the simple Drude theory. The available photoemission data also in general indicate strong damping effects and smallness of quasi-particles weight near the Fermi surface. A common origin of this incoherence may be proximity of the Mott insulator with dominance of incoherent excitations generated by disorder of spin or orbital degrees of freedom at low temperatures. In other words, spin and orbital fluctuations scatter charge excitations incoherently if they are not ordered. Among the transition metal oxides, Mn oxides, $La_{1-x}Sr_xMnO_3$ have a unique character, where ferromagnetic metals exist with a perfect spin polarization at low temperatures due to the double exchange mechanism. Even under the spin polarization, the Mn compounds show a similar or even more incoherent charge dynamics than other transition metal oxides [2, 3, 4, 5]. This implies importance of orbital fluctuations presumably coupled with lattice fluctuations, because the spin degrees of freedom are not able to contribute to the incoherence. The orbital degrees of freedom coupled to lattice are also expected to play an important role in the insulating phase through the cooperative Jahn-Teller ordering.

An important and puzzling feature in the ferromagnetic metallic phase of $La_{1-x}Sr_xMnO_3$ is that the Drude weight relative to the noninteracting system is very small ($\sim$ 0.01-0.05) while the specific heat coefficient $\gamma$ is also small ($\sim$ 2-5 mJ/moleK$^2$) [6]. The former is a typical indication of the incoherent charge dynamics, which is likely to be a consequence of orbital and lattice fluctuations. Naively, these fluctuations and resultant residual entropy are expected to increase $\gamma$, which is not the case in this compound.

In this paper, using numerical methods, we analyze roles of orbital degrees of freedom in a spin polarized two-dimensional model [7, 8]. To understand fluctuation effects in detail we employ a quantum Monte Carlo (QMC) and Lanczos diagonalization methods for numerical calculations. Our numerical results are summarized as follows: At half filling where the $e_g$ orbital is occupied by one electron per site on average, interorbital on-site Coulomb repulsions $U$ opens the Mott charge gap . The opening of the gap is, however, substantially slower than the Hartree-Fock prediction of the present model and also than the QMC results of the corresponding single-band Hubbard model. The staggered orbital order appears at the Mott insulating phase, while the order is destroyed quickly if the system is metallized by doping of carriers. The short-ranged orbital correlation shows critical enhancement when the Mott insulator is approached. This is similar to the criticality observed in the single-band Hubbard model, where spin correlations are enhanced instead of the orbital correlations. In accordance with critical fluctuations of the orbital correlations, the Drude weight is more strongly suppressed with decreasing doping concentration than the total intraband weight.

When the coupling of the orbital to Jahn-Teller distortions is introduced, the charge gap is strongly enhanced through the optimized Jahn-Teller distortion. We optimize the static Jahn-Teller distortion to lower the total energy including electronic, lattice and their coupled part. This corresponds to a classical treatment of the lattice. A reallistic choice of electron-lattice coupling suggests that the Jahn-Teller distortion and the on-site Coulomb repulsion are both important for the formation of the charge gap in $LaMnO_3$ through synergetic effects of these two. With this choice of parameters, the charge gap amplitude, stabilization energy and amplitude of the Jahn-Teller distortion are all rather quantitatively reproduced. In the metallic phase, we argue that the Jahn-Teller coupling induces an instability to phase separation near the Mott insulator. We also discuss effects of three dimensionality, which retains small Drude weight but contributes to the suppression of the specific heat coefficient $\gamma$ through an anisotropic scaling [9].

# 2 Model

We study an orbitally degenerate model of the manganese oxides with electronic, lattice and their coupled terms described by

$$\mathcal{H} = \mathcal{H}_{\rm el} + \mathcal{H}_{\rm ph} + \mathcal{H}_{\rm el-ph}. \tag{1}$$

The first term $\mathcal{H}_{\rm el}$ represents the electronic part which contains the kinetic energy and the Coulomb interaction. A $Mn^{3+}$ ion in $LaMnO_3$ has four $3d$ electrons. Three electrons occupy $t_{2g}$ orbital while the last electron occupies one of the doubly degenerate $e_g$ orbital, where all of the four electrons have parallel alignment of the spins due to strong Hund's rule coupling. In the ferromagnetic and metallic phase, the spins are perfectly polarized with the moment $\sim 3.8\mu_B$ at low temperatures and the spin degrees of freedom do not play a role. Therefore a model which ignores the spin degrees of freedom would be a good starting point to discuss low-energy physics of the Mn compounds in the metallic phase. Moreover, the Mott insulating phase has $A$-type antiferromagnetic order in which the spins are ordered in parallel in a two-dimensional (2D) plane. Therefore, a 2D plane isolated from the perovskite structure always has a ferromagnetic polarization near the Mott insulator. We will discuss later that the 3D coupling only contributes to a detailed low-energy structure in the insulating phase, as naturally expected from magnetic as well as orbital orderings with strong 2D anisotropy. Although the orbital long-ranged order is lost, the strong 2D anisotropy may also be retained in metals, and we study the metal-insulator transition in the 2D model as a starting point. We later discuss 3D effects on the metal-insulator transition based on these 2D results. We also note that our model may have some relevance to layered Mn compounds [10], which consist of stacking of single or double $MnO_2$ layers with rather small crystal-field splitting of $d_{x^2-y^2}$ and $d_{3z^2-r^2}$. All of the above problems may be studied in an isolated 2D plane at least as a good starting point. We employ such an appropriate 2D model by ignoring the spin degrees of freedom. In this model, we still have orbital degrees of freedom for $e_g$ electrons. Therefore the electronic part of the Hamiltonian $\mathcal{H}_{\rm el}$ describes $e_g$ electrons with double degeneracy as

$$\begin{aligned}\mathcal{H}_{\rm el} &= -\sum_{\langle ij\rangle,\nu,\nu'} t_{i,j,\nu,\nu'}(c^\dagger_{i\nu}c_{j\nu'}+{\rm H.c.})\\ &\quad +U\sum_i(n_{i1}-\frac{1}{2})(n_{i2}-\frac{1}{2})-\mu\sum_{i\nu}n_{i\nu},\end{aligned}\tag{2}$$

where the creation (annihilation) of the single-band electron at site $i$ with orbital $\nu$ is denoted by $c^\dagger_{i\nu}(c_{i\nu})$ with $n_{i\nu}$ being the number operator $n_{i\nu}\equiv c^\dagger_{i\nu}c_{i\nu}$. The Hamiltonian looks similar to the ordinary Hubbard model if the orbitals are replaced with spins. However, an important difference from the ordinary Hubbard model is that the transfer has an anisotropy and off-diagonal elements with dependence on the $d_{x^2-y^2}$ and $d_{3z^2-r^2}$ orbitals as well as on spatial directions, where we label the orbital $d_{x^2-y^2}$ and $d_{3z^2-r^2}$ as $\nu=1$ and 2, respectively below. In 2D configuration, the nearest-neighbor transfer is scaled by a single parameter $t_0$ as $t_{11}=\frac{3}{4}t_0,\ t_{22}=\frac{1}{4}t_0,\ t_{12}=t_{21}=\pm\frac{\sqrt{3}}{4}t_0$, where, in $\pm$, $+(-)$ is for the transfer to the $y(x)$ direction. We note that the interaction $U$ is not a bare interorbital Coulomb repulsion $U_{12}$ but $U_{12}-J_{12}$, where $J_{12}$ is the Hund's rule coupling between orbitals $\nu=1$ and 2. In this paper we take the energy unit as $t_0$.

In the absence of $U$, the diagonalized two bands have dispersions both with the bandwidth $4t_0$, where the energy of one band is $2t_0$ higher as a whole than the other. We have to note that the perfect nesting condition is satisfied at half filling in the noninteracting isolated 2D plane although it is not in the real 3D structure. In the 2D lattice at half filling, the lower band is at three-quarters filling while the upper band is at a quarter filling, both of which have square shapes of the Fermi surfaces. The nesting vector $(\pi,\pi)$ connects two Fermi surfaces belonging to different bands. Because of this nesting condition in the 2D lattice, it may have stronger tendency to the Mott insulator with the orbital order at half filling than the 3D structure.

The electronic degrees of freedom are coupled with lattice through several different types of distortions [11]. We consider the Jahn-Teller coupling in the Hamiltonian (1), because it linearly and strongly couples with the $e_g$ orbital fluctuations and ordering, which is the main subject of this paper. The Jahn-Teller distortion coupled to two $e_g$ orbitals is represented by two orthogonal modes of oxygen displacements $u_i$ at the octahedron corner surrounding a Mn ion at the $i$-th site. A linear combination of these two modes $Q_2$ and $Q_3$ may represent any Jahn-Teller-type distortion coupled to the $e_g$ orbitals, expanded as a two-component vector as $\mathbf{v}_i = -2u_i(\cos 2\theta_i, \sin 2\theta_i)$, where $\theta_i$ determines the angle of the linear combination. The $Q_2$ and $Q_3$ modes linearly couple with the orbital polarization $T_i^x$ and $T_i^z$, respectively, where the pseudospin operator $T_i^\mu(\mu = x, y, z)$ is defined by $T_i^\mu = \frac{1}{2}\sum_{\nu\nu'} \hat{\sigma}^\mu_{\nu\nu'} c^\dagger_{i\nu} c_{i\nu'}$, with $\hat{\sigma}$ being the Pauli matrix. Then the electron-lattice coupling and the lattice elastic energy are given as

$$\mathcal{H}_{\rm el-ph} = \sum_i \mathbf{g}_i \cdot \mathbf{I}_i, \quad \mathcal{H}_{\rm ph} = k \sum_i u_i^2, \tag{3}$$

where $\mathbf{g}_i = g\mathbf{v}_i$ and $\mathbf{I}_i = (T_i^z, T_i^x)$. The electron-lattice coupling constant and the spring constant for the Jahn-Teller lattice distortion are represented by $g$ and $k$, respectively. We treat $u_i$ and $\theta_i$ as a classical variable [12]. In the following the possibility of static and uniform Jahn-Teller distortions $u = u_i$ with experimentally-observed staggered pattern $\theta_i = \frac{\pi}{6}(-1)^{|\mathbf{r}_{ix}|+|\mathbf{r}_{iy}|}$ is considered by minimizing the total energy.

When the Jahn-Teller distortion is nonzero, the transfer amplitude $t_0$ is also modified through the oxygen distortion. Empirically it is expected that the Mn-O transfer changes by a factor $(1 \pm 2u_i)^{-7/2}$ for the oxygen distortion $u_i$ [13]. In this paper, we take the length unit as the Mn-Mn distance. Then, although this has minor effects on our results shown below, we take the modified Mn-Mn transfer as $\tilde{t}_0 = t_0(1 - 4u_i^2)^{-7/2}$.

The model (1) is suited to understand two different problems. The first is the basic electronic structure where orbital-lattice coupled systems primarily determines the insulating charge gap while the intersite spin correlations play secondary roles. The second is the incoherent charge dynamics in metals at low temperatures where spin degrees of freedom are indeed absent due to the perfect polarization.

We employ the projector quantum Monte Carlo (PQMC) [14, 15] and the Lanczos diagonalization (LD) methods for the model (1)-(3) to calculate physical properties in the ground state. In the PQMC, the minus sign problem poses a restriction to tractable range of the parameters. We show the cases in which convergence to the ground state is reached without this problem and the ground state properties are obtained with needed accuracy. In the next section, the charge gap and orbital order and Jahn-Teller distortion are presented after taking the extrapolation to the infinite system size (thermodynamic limit). In the LD, system size is limited by the computer memory size. We calculate two sizes, 4 by 4 and $\sqrt{10}$ by $\sqrt{10}$ lattices. The boundary condition is chosen so as to realize the ground state, where phase difference $\phi$ is varied in $x$ and $y$ directions. Note that $\phi = 0$ and $\phi = \pi$ correspond to the periodic and antiperiodic boundary conditions, respectively.

As realistic choices of parameters in the Hamiltonian, we take the following [4, 16, 17, 18]: The effective Mn-Mn transfer $t_0$ is estimated around 0.5eV while the on-site Coulomb repulsion is varied in the range of $0 \leq U/t_0 \leq 16$ to see the $U$ dependence. The elastic constant $k$ is roughly estimated from the Mn-O stretching phonon frequency ($\sim$ 70 meV) as several tens eV in the length unit of Mn-Mn distance [2, 3, 12]. In our calculation, we take $k/t_0 = 100$. In the same length unit, the Jahn-Teller coupling $g$ is taken as $g/t_0 = 10$, which can be roughly estimated from the electrostatic energy between an $O^{2-}$ ion and an electron on different types of $3d$ orbital on the neighboring Mn site.

# 3 Results

The results are discussed in the following order: First we discuss results at half filling. The ground state properties of $\mathcal{H}_{\rm el}$ without the Jahn-Teller coupling are presented. We next show how the results are modified when we introduce the Jahn-Teller coupling. In the second part, carriers are doped and the metallic phase is studied. Away from half filling, we mainly show properties in the absence of the Jahn-Teller coupling, although its effects are briefly discussed in the next section.

At half filling without the Jahn-Teller coupling, the charge gap and the orbital order were calculated by the PQMC method. As we see in Fig. 1, the charge gap amplitude $\Delta_c$ is substantially smaller than the Hartree-Fock prediction. Although the nesting condition is satisfied in the 2D model, the growth of the charge gap with increasing $U$ is slow. This slow growth is also confirmed from the comparison with the PQMC results of the ordinary single-band Hubbard model. The charge gap at $U/t_0 = 4$ is around 0.66 for the single-band Hubbard model while it is around 0.1, in the present case. To reproduce the Mott charge gap in the experimental result of $LaMnO_3$, $\Delta_c \sim 1$eV [19], we need much larger $U$. This may also be true in real 3D structure because the Mott gap opening would be even slower due to the absence of the perfect nesting. The slow growth of the gap may be due to fluctuations and competition between interband off-diagonal and intraband hoppings which both satisfy the perfect nesting condition while the orbital order is stabilized only through the interband part. The orbital order is also plotted in Fig. 1, where the order grows quickly beyond some threshold $U/t_0 \sim 4$. We note that this orbital polarization has a uniaxial symmetry.

When the Jahn-Teller coupling is introduced with static and uniform Jahn-Teller distortion $u$ with the staggered pattern $\theta_i$ being allowed, the ground state is achieved at a nonzero $u$. Figure 2 shows the PQMC results of the ground state energy at $U/t_0 = 4$ as function of static and uniform distortion $u$. With this choice of parameters $k$ and $g$, the $U$ dependence of the optimized distortion is plotted in Fig. 3. The experimentally observed value $u = 0.035$ for $LaMnO_3$ is close to the values of the optimized distortions plotted in this figure. Note that the optimized distortion $u$ at $U = 0$ is nearly zero. The small deviation between our numerical results around $U = 4$ and the experimental results would possibly be due to the uncertainty in the choice of $k$ and $g$. We will discuss this point later. The $U$ dependence of the charge gap at the optimized distortion is plotted in Fig. 3. As compared to the case $g = 0$, opening of the charge gap is much quicker with increasing $U$. We note that the gap $\Delta_c$ at $U = 0$ is around $0.02t_0$ and far below the experimentally observed value $\sim 2t_0$. The quick opening of the gap with increasing $U$ comes from synergetic and nonlinear effects between electron correlations $U$ and the Jahn-Teller distortions. The Jahn-Teller distortions and the on-site Coulomb repulsion cooperatively stabilize and enhance the charge gap of the insulating state through the dynamic coupling of orbital and lattice polarizations. The experimentally observed charge gap $\Delta_c \sim 1$eV ($\Delta_c/t_0 \sim 2$) is numerically reproduced if we take $U/t_0$ at around 4 and 5.

The ground state energy difference, $E_{JT}$, between the case with optimized Jahn-Teller distortion and that at $u = 0$ roughly measures the stabilization energy of the Jahn-Teller distortion. This is around 0.08eV ($\sim 0.1$-$0.2t_0$) if we take $U/t_0 = 4$-$5$ as seen in Fig. 3. It is again favorably compared with the Jahn-Teller transition temperature $\sim 800$K.

From these analyses, a consistent picture to explain the experimental and the present numerical results emerges : (1) The charge gap of the order of 1eV is determined by a synergetic effect of the on-site Coulomb repulsion and the atomic Jahn-Teller distortion. (2) The Jahn-Teller transition temperature accompanied by the orbital long-ranged order is roughly determined from intersite coupling for combined Jahn-Teller and orbital exchange. The intersite orbital correlation and intersite correlations of the Jahn-Teller distortion are both developed in the energy scale of $\sim 0.1$eV. (3) The interlayer antiferromagnetic spin correlation grows at much lower energy scale around 0.02 eV.

In our analyses, although we have fully taken into account the orbital correlations including short-ranged and dynamical fluctuations, we have not considered dynamical effects for the Jahn-Teller distortions. This will have the following effects: When a hole is doped, the Jahn-Teller distortion may relax around the hole site to form a polaron. The one-hole energy then decreases and the charge gap $\Delta_{\rm c}$ may thus to some extent be reduced from the present value. Next, because $E_{\rm JT}$ is the energy difference between the Jahn-Teller distorted and undistorted phases, it may overestimate the energy scale of the Jahn-Teller transition temperature $T_{\rm JT}$ because $E_{\rm JT}$ contains not only the intersite Jahn-Teller ordering energy but also the on-site Jahn-Teller polarization energy, while $T_{\rm JT}$ is determined basically from the intersite part [11]. Presumably, larger values of $k$ and $g$ than the present choices would be required to reproduce $\Delta_{\rm c}$ and $T_{\rm JT}$ if we consider the short-ranged Jahn-Teller correlation effects. We need further studies on this short-ranged-correlation effects and also on the uncertainties in the choice of the parameter values for $k$ and $g$.

We next discuss the doped cases. When the Jahn-Teller coupling is absent, the orbital order clearly observed in the half-filled Mott insulator is quickly destroyed with increasing doping concentration $\delta$. In Fig. 4, equal-time structure factor of the orbital correlation $T^x(\mathbf{Q})$ is plotted at $\mathbf{Q} = (\pi, \pi)$ as function of hole concentration $\delta$. Here we define $T^\mu(\mathbf{k}) = N_{\rm S}^{-1} \sum_{ij} \langle T_i^\mu T_j^\mu \rangle \exp(\mathrm{i}\mathbf{k}\mathbf{r}_{ij})$ $(\mu = x, y, z)$ for an $N_{\rm s}$-site system. This indicates that $T^x(\mathbf{Q})$ is scaled by $\delta^{-1}$ hence the orbital correlation length is scaled by the mean hole distance $\delta^{-1/2}$. This is the same scaling as the antiferromagnetic correlation in the 2D single-band Hubbard model [20]. The doping concentration dependence of the chemical potential is plotted in Fig. 5. This implies that the chemical potential is proportional to $\delta^2$, which results in the scaling of the charge susceptibility $\chi_c \equiv dn/d\mu \propto \delta^{-1}$. This is again the same scaling as the 2D single-band Hubbard model [20, 21].

These suggest that the critical enhancement of the orbital correlation plays a similar role to spin correlations in the single-band Hubbard model and the metal-insulator transition is controlled by a similar criticality if the Jahn-Teller coupling is switched off. Although it is not numerically confirmed, it is tempting to ascribe the enhancement of the charge susceptibility to flat dispersions generated at some particular points of the momentum space in the single-particle excitation near the Fermi level at small $\delta$, as is indeed observed in the single-band Hubbard model [22]. This may generate dynamical fluctuations for the phase separation or dynamical charge-order fluctuations related to the location of the flat dispersions.

The optical conductivity calculated by the LD method shows reduction of the Drude weight with decreasing $\delta$. This reduced Drude response coexists with a conspicuous intraband incoherent response at finite frequencies and also transition to the upper (or lower) Hubbard band at high frequencies as seen in Fig. 6. In this figure, $U$ is taken larger than the realistic values to separate the intraband incoherent response from the inter-Hubbard-band transition. In the experimental situation, these two finite-frequency weights are merged to a single structure. A large incoherent response at finite $\omega$ within the Mott gap may be a consequence of carrier scattering by short-ranged orbital correlations. The optical conductivity will be studied elsewhere in more detail [8].

# 4 Discussion

Quantum Monte Carlo results indicate that for realistic choice of the parameters of $LaMnO_3$, the insulating charge gap ($\sim$ 1eV) at half filling is a consequence of synergetic and cooperative enhancement by the "Mott gap" and "Jahn-Teller gap". The energy scale of intersite correlations for the Jahn-Teller distortions and the orbital correlations is one order of magnitude smaller than the charge gap energy scale ($\sim$ 1eV) . The energy scales of antiferromagnetic correlations and the three-dimensional effects are two orders of magnitude smaller than the charge gap scale. The experimental results for the charge gap $\Delta_c$ and the Jahn-Teller distortion $u$ are favorably compared with our results at $U/t_0 = 4$-$5, g/t_0 \sim 10, k/t_0 \sim 100$ and $t_0 \sim 0.5$eV. To make more precise comparisons possible between numerical and experimental values, a more scrutinized tuning of the parameters would be required. In the absence of the Jahn-Teller coupling, the orbital long-ranged order stabilized at half filling is destroyed quickly upon doping while the short-ranged orbital correlation is critically enhanced with $\delta \to 0$ in metals. The charge susceptibility shows critical enhancement with $\delta \to 0$ implying the formation of a flat dispersion at some particular points in the momentum space of the single-particle excitation.

When the Jahn-Teller coupling is introduced for the doped cases, our numerical results suggest that, in the absence of long-ranged Coulomb interaction, the phase separation may take place in a finite range of doping concentration typically between $\delta = 0$ and $\delta \simeq 0.2$. This phase separation is resulted from a combined effect of the already enhanced charge susceptibility at $g = 0$ and the Jahn-Teller effect. Note that the phase separation takes place if the charge susceptibility diverges. A similar result of the phase separation was suggested in a study of a one-dimensional model [23]. With longer-ranged Coulomb interaction, charge ordering may take place instead of real phase separation. The Jahn-Teller coupling promotes such tendency toward real-space structure in non-half-filled systems. This problem will be discussed elsewhere in detail [24].

If a dynamical charge fluctuation and flat dispersions appear, strong reduction of the Drude weight is a natural consequence as already clarified in the studies of the 2D $t$-$J$ model [25]. The scaling theory of 2D system also predicts critical enhancement of the specific heat coefficient $\gamma$ [26]. This is however, experimentally not observed as discussed in §1. This problem was examined in terms of three-dimensional effects [9]. When a flat dispersion in the form of $\sim k^4$ appears in the intraplane direction while the dispersion is normal in the interlayer direction, the specific heat coefficient follows $\gamma \propto \delta^0$ and does not show critical enhancement while the Drude weight follows $D \propto \delta^{4/3}$. This seems to be consistent with the experimental observation. We need further studies to fully understand the microscopic basis in detail. The momentum dependence of the single-particle excitations is an important issue to be explored.

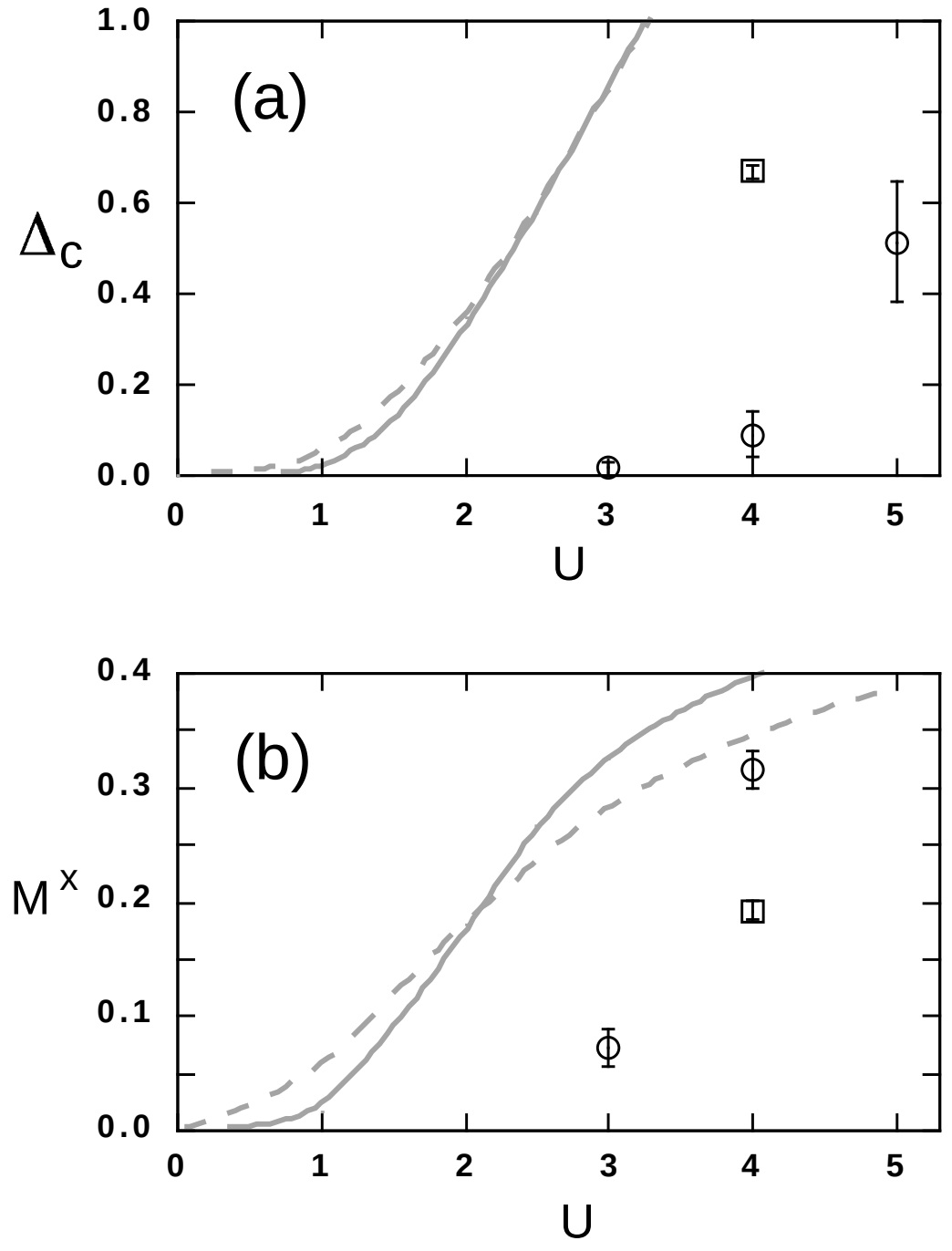

Figure 1: QMC results (circles) without the Jahn-Teller coupling for $U$ dependences of (a) the charge gap and (b) the staggered orbital polarization. The data are shown in the limit of $N_{\rm S} \to \infty$, which are obtained from size extrapolations of the data on the sizes $N_{\rm S} = 4 \times 4$ to $10 \times 10$. Squares are QMC results for the ordinary single-band Hubbard models. The gray (dotted) curves show the Hartree-Fock results for our model (for the ordinary single-band Hubbard model).

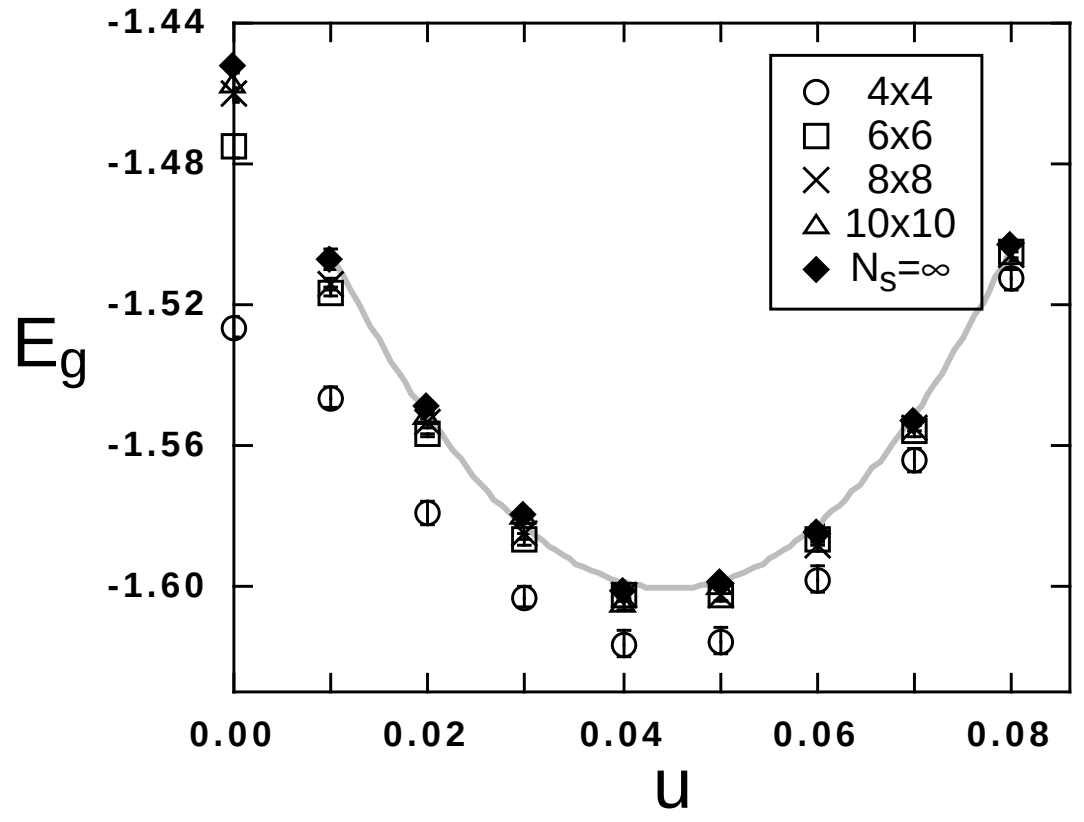

Figure 2: Ground state energy per site at half filling as function of the oxygen distortion $u$. We take $U/t_0 = 4$, $g/t_0 = 10$ and $k/t_0 = 100$. The curve is a polynomial fit to the extrapolated data to the limit $N_{\rm S} = \infty$.

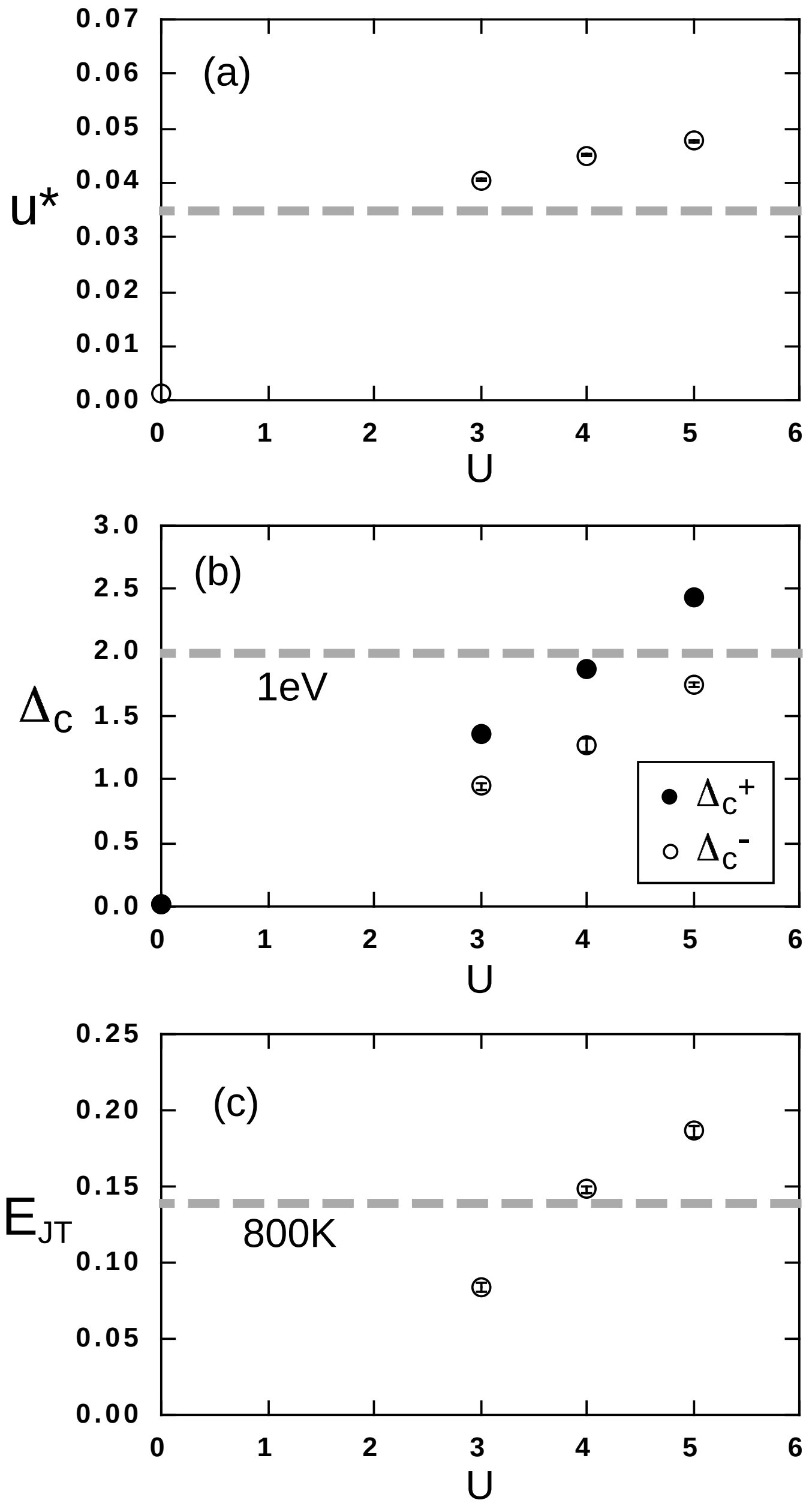

Figure 3: The $U$ dependence of (a) Jahn-Teller distortion $u^*$, for the ground state at the optimized value of $u = u^*$, (b) charge gap $\Delta_\mathrm{c}$ and (c) stabilization energy $E_\mathrm{JT}$. In (b), the charge gap is plotted for electron doping ($\Delta_\mathrm{c}^+$) and hole doping ($\Delta_\mathrm{c}^-$). Dashed lines show comparisons with experimental indications.

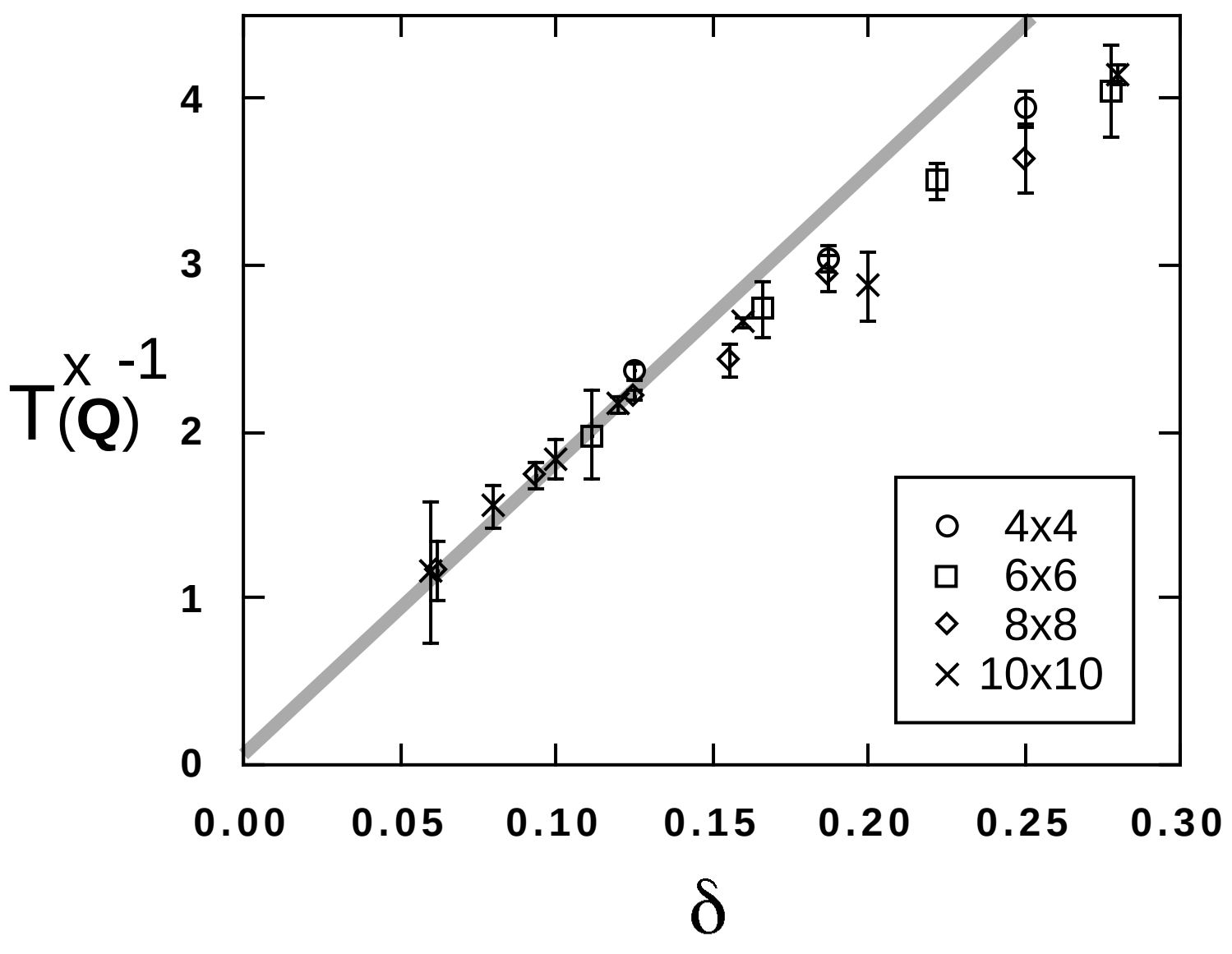

Figure 4: Inverse plot of peak values of the equal-time structure factor for the orbital correlation at the wavenumber $(\pi, \pi)$ as function of doping concentration at $U/t_0 = 4$. The gray line is the linear fit to the data at $0 < \delta < 0.15$.

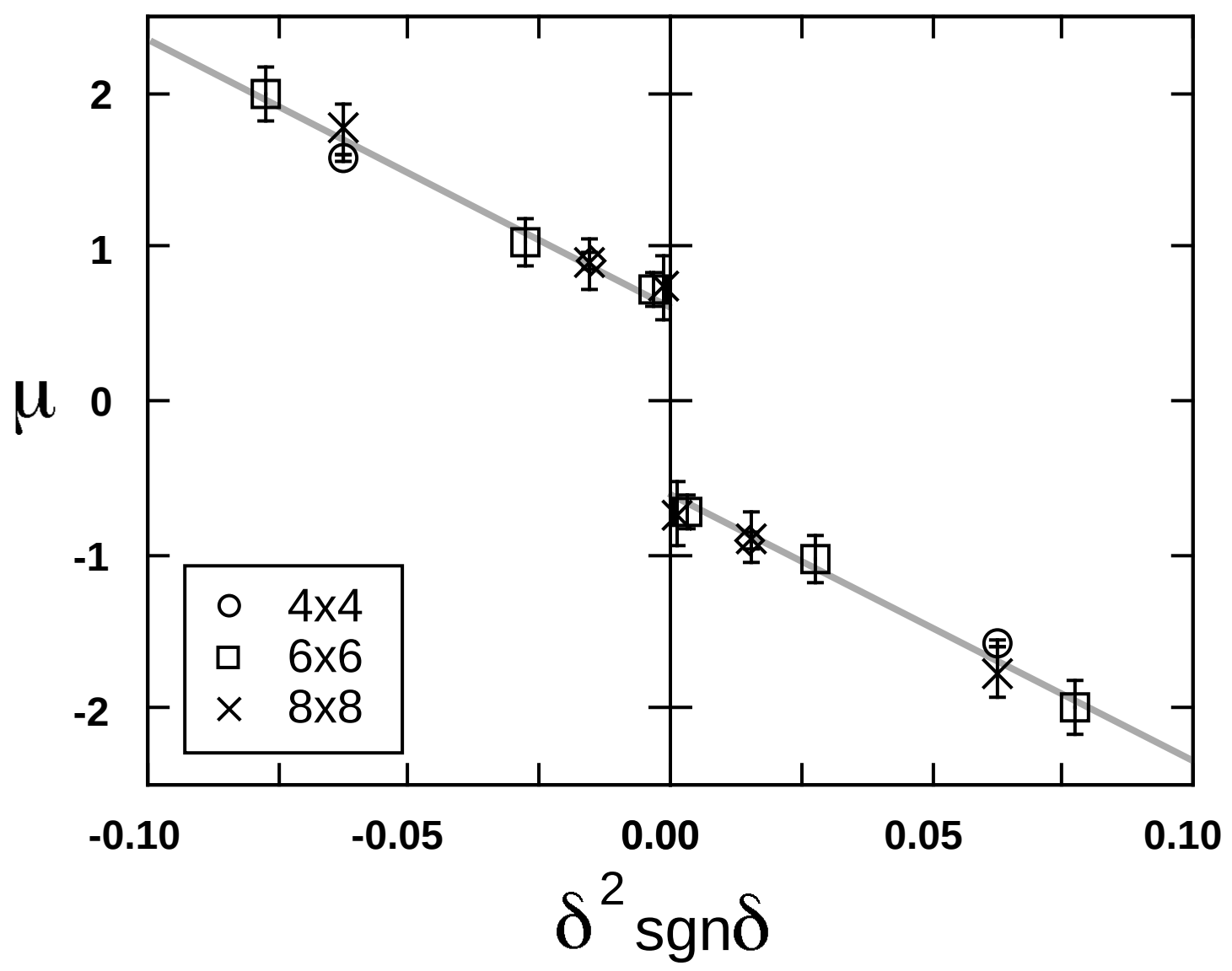

Figure 5: Doping concentration dependence of chemical potential $\mu$ at $U/t_0 = 4$. The chemical potential is scaled by $\delta^2$.

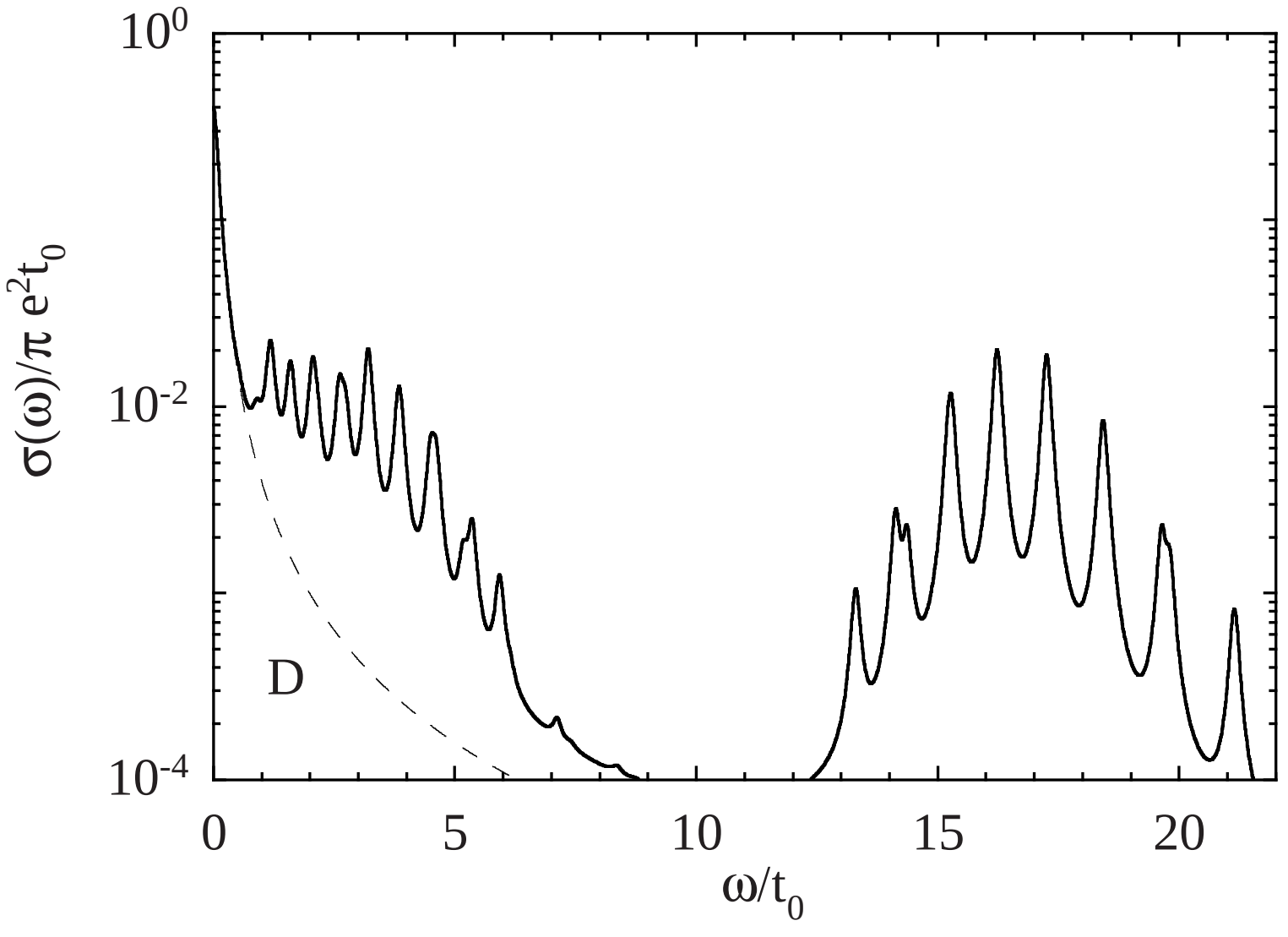

Figure 6: Optical conductivity of the Mn model at $U/t_0 = 16$ in the Lanczos diagonalization result of 14 electrons on the 4 by 4 lattice. The ground states are chosen by taking optimized phase shift of the boundary condition in $x$ and $y$ directions. The dashed curve $D$ represents the Drude part. For the convenience of illustration, the delta functions obtained in the Lanczos diagonalizations are replaced with the Lorenzians with the width $0.1t_0$.